\documentclass[namedreferences]{solarphysics}
\usepackage[optionalrh]{spr-sola-addons} 
\usepackage{graphicx}        
\usepackage{color}           
\usepackage{url}             

\newcommand{\ms}{\rm{\,m\,s^{-1}}}
\newcommand{\kms}{\rm{\,km\,s^{-1}}}
\newcommand{\Mm}{\rm{\,Mm}}
\newcommand{\s}{\rm{\,s}}
\newcommand{\secs}{\rm{\,seconds}}
\newcommand{\mHz}{\rm{\,mHz}}

\begin{document}

\begin{article}

\begin{opening}

\title{Subsurface Supergranular Vertical Flows as Measured Using
Large Distance Separations in Time--Distance Helioseismology}

\author{T.L.~\surname{Duvall Jr.}$^{1}$\sep
	  S.M.~\surname{Hanasoge}$^{2,3}$
	}
\runningauthor{T.L.Duvall Jr.,S.M.Hanasoge}
\runningtitle{Vertical Flows}

\institute{$^{1}$ Solar Physics Laboratory, NASA Goddard Space Flight Center,
Greenbelt, MD, USA 20771
email:~\url{Thomas.L.Duvall@nasa.gov}\\
$^{2}$ Max-Planck-Institut fur Sonnensystemforschung, Max Planck Stra${\beta}$e 2, 37191 Katlenburg-Lindau, Germany
email:~\url{hanasoge@mps.mpg.de}\\
$^{3}$Department of Geosciences, Princeton University, Princeton, NJ 08544, USA
}

\begin{abstract}
As large--distance rays (say, $10$\,--\,$24 ^\circ$) approach the solar surface 
approximately vertically, travel times measured from surface pairs for
these large separations are mostly sensitive to vertical flows, at
least for shallow flows within a few Mm of the solar surface.  All 
previous analyses of supergranulation have used smaller separations 
and have been hampered by the difficulty of separating the horizontal
and vertical flow components.  We find that the large separation travel
times associated with supergranulation cannot be studied using the
standard phase--speed filters of time--distance helioseismology.  These
filters, whose use is based upon a refractive model of the perturbations,
reduce the resultant travel time signal by at least an order of magnitude
at some distances.  More effective filters are derived.  Modeling suggests
that the center--annulus travel time difference 
$[\delta t_{\rm{oi}}]$
in the separation range $\Delta=10$\,--\,$24 ^\circ$ is insensitive to the 
horizontally diverging flow from the centers
of the supergranules and should lead to a constant signal from the vertical
flow.  Our measurement of this quantity, $5.1 \pm 0.1\secs$, is constant
over the distance range.  This magnitude of signal cannot be caused by
the level of upflow at cell centers seen at the photosphere of $10\ms$ 
extended in depth.  It requires the vertical flow to increase with depth.
A simple Gaussian model of the increase with depth implies a peak upward 
flow of $240\ms$ at a depth of $2.3\Mm$ and a peak horizontal flow of
$700\ms$ at a depth of $1.6\Mm$.  
\end{abstract}
\end{opening}

\section{Introduction}
     \label{S-Introduction} 

Supergranulation, first studied more than 50 years ago 
\cite{Hart54,Leighton62}, continues
to be an active area of research.  A comprehensive review \cite{Rieutord10}
details the recent progress.  
It has proven difficult to measure the vertical flow of supergranules
at the photospheric level.  The recent measurements of a $10\ms$ upflow
at the center of the average cell with a horizontal variation consistent with
a simple convection cell \cite{Duvall10} may have finally settled the
issue.  It is only through helioseismology that we would hope to measure
the supergranulation flows below the photosphere.  \inlinecite{Duvall97}
first showed that time--distance helioseismic techniques are sensitive
to supergranules and that inversions to derive three--dimensional flow
fields might be derived.  However, \inlinecite{Zhao03} showed that their 
ray--theory
inversions could not separate the horizontal and vertical flows for a
model flow field.  The use of radial--order  filters and Born 
approximation kernels
has led to more successful separation \cite{Jackiewicz08,Svanda11}.  
\inlinecite{Gizon10} summarize the local helioseismic contributions to
supergranulation.

An important quantity in time--distance helioseisomology is the 
arc separation $[\Delta]$ between pairs of photospheric locations 
whose signals are subsequently temporally cross correlated.
Previous analyses used
$\Delta$ less than $5^\circ$ ($61\Mm$) \cite{Zhao03} or
$2.4^\circ$ ($29.2\Mm$) \cite{Jackiewicz08,Svanda11}, as the 
sensitivity is greater for the horizontal flow, 
the signal--to--noise ratio is better for small separations, 
and basically no supergranulation signal could be seen at larger $\Delta$ in 
the travel time difference maps constructed from the relatively short 
(8--12 hours) time intervals required to study the one--day lifetime 
supergranules.  In the present work, we show that the large separations
of $\Delta=10$\,--\,$24^\circ$ yield center--annulus travel time differences 
$[\delta t_{\rm{oi}}\equiv$ outward--going time minus inward--going time$]$ from the
centers of average supergranules that are insensitive to the diverging
horizontal flow and hence yield a purely vertical flow response.

In Section~\ref{sec:1} the modeling efforts, including 
flow models that satisfy mass conservation, travel times computed from
ray theory, and linear wave simulations through flow fields are presented.
In Section~\ref{sec:danal}, the analysis of data is presented, and in
Section~\ref{sec:discuss} the results are discussed.

\section{Modeling}
\label{sec:1}
\subsection{Flow Model}
\label{sec:2.1}
As our observational technique is to make averages about centers of
supergranules, the flow model that we take is azimuthally symmetric
and decays exponentially away from the cell center
to simulate the effect of averaging a stochastic flow field.  A time--independent
flow is assumed with axisymmetry about the location of the average cell.
A Cartesian coordinate system with $z$ positive upwards is defined.
For these assumptions, the vanishing divergence of the mass flux
implies
\begin{equation}
\frac{\partial}{\partial z} (\rho v_z) = -\rho\bf{\nabla_h}\cdot\bf{v_h},
\label{eq:cont}
\end{equation}
where $\rho$ is the $z$-dependent density, $v_z$ is the $z$-component of
velocity, $\bf{\nabla_h}$ is the horizontal divergence, and $\bf{v_h}$ is
the horizontal velocity.  
It is assumed that horizontal density variations can be ignored,
as travel times are only weakly sensitive to density changes in the
underlying medium \cite{Hanasoge12}. 
The model is assumed to separate into horizontal
and vertical functions
\begin{equation}
{\bf{v_h}} \equiv -f(z) {\bf{g}}(x,y),
\end{equation}
\begin{equation}
v_z\equiv u(z)\bf{\nabla_h}\cdot\bf{g},
\end{equation}
where ${\bf{g}}(x,y)$ is a vector in the horizontal plane with no units,
$f(z)$ has units of velocity, and $u(z)$ has units of velocity times
length.
Our method of solution is to first specify the horizontal function
${\bf g}(x,y)$ and calculate its horizontal divergence 
$\bf{\nabla_h}\cdot\bf{g}$.  The vertical function $u(z)$ is then specified, and
Equation~(\ref{eq:cont}) is used to derive $f(z)$.
Some straightforward algebra yields
\begin{equation}
f(z) = \frac{\partial u}{\partial z} +u(z)\frac{\partial \ln \rho}{\partial z}.
\end{equation}
In general, models are considered with the horizontal 
function ${\bf{g}}(x,y)$ defined by
\begin{equation}
{\bf{g}}(x,y) = {\bf \hat{r}}J_1(kr)e^{-r/R},
\label{eq:gdef}
\end{equation}
where ${\bf \hat{r}}$ is the outward radial unit vector in a cylindrical
coordinate system, $J_1$ is the Bessel function of order one, $k$ is a
wavenumber, $r$ is the horizontal distance from the origin, and $R$
is a decay length.  
$k$ and $R$ are, in principle, free parameters that will only have 
the default values
$k=\frac{2\pi}{30} \rm{\,rad\,Mm^{-1}}$ and $R=15\Mm$ in this article.
This type of horizontal variation was used by 
\inlinecite{Birch07}.
This type of horizontal variation describes well the type of averaging of
the supergranular field done in this paper, where one sees the outward
flow as the first positive lobe of the $J_1$ function and on average,
the inflow from the adjacent supergranular cells as the first negative 
range of the $J_1$ function.


In general, models have been considered with a Gaussian $z$-dependence.
$v_z$ is specified at $r=0$ as a simple Gaussian:
\begin{equation}
v_z(r=0) = k u(z) = v_0 e^{-(z-z_0)^2/2\sigma_z^2},
\end{equation}
where $z_0$ is the location of the peak of the vertical flow, $\sigma_z$ is the
Gaussian sigma, and $v_0$ is the maximum vertical flow.  To ultimately 
explain the large--distance travel times, the photosphere needs to be in
the far tail of the Gaussian.  As the upward flow at cell center is
$10\ms$ at the photosphere, this implies a considerably larger 
vertical flow at depth for
the average cell.  Values for these parameters that approximate the
data are $v_0=240\ms$, $z_0=-2.3\Mm$, and $\sigma_z=0.912\Mm$.

\subsection{Ray Calculations}
\label{sec:rays}

In the ray theory, the travel time difference for the two directions 
of propagation through
a flow $\bf v$ is
\begin{equation}
{\delta\tau}=-2\int_{\Gamma}\frac{{\bf v}\cdot{\bf{ds}}}{c^2},
\label{eq:tdif}
\end{equation}
where $\Gamma$ is the ray path, $c$ is the sound speed, 
and ${\bf{ds}}$ is the element of length in
\begin{figure}
\centerline{\includegraphics[width=1.0\textwidth,clip=]{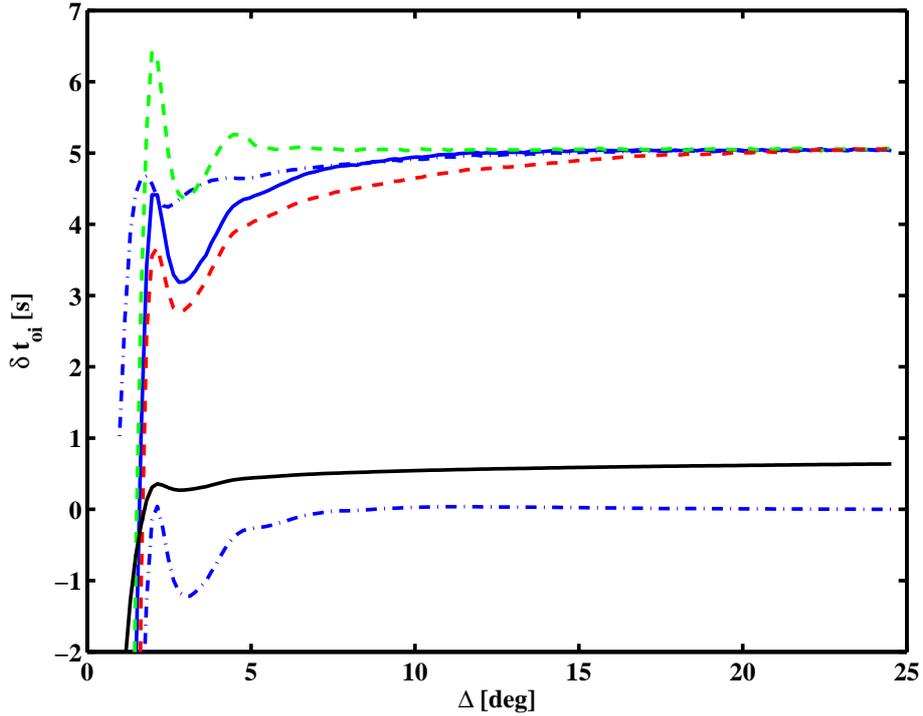}
           }
\caption{
Comparison of various ray models.  The blue curves are for the Gaussian 
model with $z_0=-2.3\Mm$.  The lower dot--dashed blue curve is for the 
horizontal
flow component, the upper dot--dashed blue curve is for the vertical 
flow component and
the solid blue curve is for the sum (upper solid curve).  
The upper dashed (green) curve is for the sum for
$z_0=-1.15\Mm$ model and the lower dashed (red) curve is for the sum of 
the $z_0=-3.45\Mm$
model.  The (lower) solid black curve is the sum for a model with a constant upflow
of $10\ms$.
}
\label{F-model1}
\end{figure}
the direction of the ray \cite{Giles00}.  This equation naturally separates
into two terms for the horizontal and vertical flow contributions to the
travel time.  
The ray generation and raw ray integrations are performed using
code developed and discussed in detail by \inlinecite{Giles00}.
This code was extended to integrate quadrant and annulus surface--focusing
integrations for 3D flow and sound speed perturbation models.
In Figure~\ref{F-model1} the black curve shows the travel
time difference $[\delta t_{\rm{oi}}]$ for a flow model with a 
constant upflow of $10\ms$.  The
black curve is for the sum of horizontal and vertical flows.  That is, 
a horizontal flow of the form of Equation~(\ref{eq:gdef}) consistent 
through the continuity equation 
with a $10\ms$ upflow is used.  This signal, with a maximum of $0.6\secs$,
is much too small to explain the results derived 
in Section~\ref{sec:gamma}.  This implies that
the vertical flow must increase with depth from the photospheric value
of $10\ms$.

Also in Figure~\ref{F-model1}, the separate horizontal and vertical 
travel time contributions
and the sum are shown for the Gaussian model mentioned above.  For this 
relatively shallow model, 
it is seen that in the distance range $10$\,--\,$24^\circ$, the signal is essentially
constant and is mostly due to the vertical flow.  Towards shorter distances,
the horizontal flow contributes an increasing amount.  The Gaussian model has
three free parameters: $v_0$, $z_0$, and $\sigma_z$.  In a later section the
travel time difference for $10$\,--\,$24^\circ$ will be determined to be 
$5.1\s$.  This
travel time difference and the $10\ms$ upflow at the photosphere serve to
determine two of the three free parameters.  
The travel time at $\Delta=24.5^\circ$ is forced to $5.1\secs$.
Letting $z_0$ be the free
parameter, three models are plotted in Figure~\ref{F-model1} with values
of $z_0=-1.15\Mm$, $z_0=-2.30\Mm$, and $z_0=-3.45\Mm$.
The shallowest model shows the least contribution from the horizontal 
component in the distance range $5$\,--\,$10^\circ$.  This distance range may 
supply a way to distinguish the best model while comparing with data.

A summary of the characteristics of these models is contained in 
Table~\ref{T-models}.  The maximum vertical flow is at least a factor of 20
larger than the photospheric value.  The ratio of the maximum 
horizontal flow
to the photospheric value varies between four and eight.  The maxima of the 
horizontal flow occurs somewhat shallower than the peak vertical flow.

\begin{table}
\caption{ 
Model details.  The three models g1, g2, g3 are ones with Gaussian (g) upward
flow at cell center peaking at $z0\Mm$ with peak flow $v_z(z0)\ms$ and with
Gaussian width $\sigma_z\Mm$.  The upward flow at the top (z=0) is
$v_z(0)\ms$.  The peak horizontal flow at the top is $v_h^{\rm{max}}(0)\ms$ 
which occurs at radial distance $r^{\rm{\rm{max}}}\Mm$.  The peak outward flow 
$v_h^{\rm{max}}\ms$ occurs at $z(v_h^{\rm{max}})\Mm$.  
All of the maxima of $v_h$ occur
at the same $r^{\rm{max}}$ by construction.  The model sim1 is the one for the
small features with just a z flow that are used for the flow perturbation 
described in Section~\ref{sec:sim}.
}
\label{T-models}
\begin{tabular}{ccccccccc}
\hline
Name & $z_0$ & $\sigma_z$ & $v_z(0)$ & $v_h^{\rm{max}}(0)$ & $r^{\rm{max}}$ & $v_z(z0)$ & $v_h^{\rm{max}}$ & $z(v_h^{\rm{max}})$ \\
\hline
g1   & -3.45    & 1.39          & 10       & 122     & 7 & 218 & 460 & -2.44\\
g2   & -2.3     & 0.912         & 10       & 138     & 7 & 240 & 697 & -1.62\\
g3   & -1.15    & 0.433         & 10       & 195     & 7 & 340 & 1609 & -0.83\\
sim1 & -2.3     & 0.816         & 6.3      &         &   & 338 &      &\\
\hline
\end{tabular}
\end{table}

\subsection{Simulation Results}
\label{sec:sim}
The applicability of ray theory to time--distance helioseismology has been
called into question \cite{Bogdan97}.  There are cases
for supergranulation studies (\opencite{Birch07}, Figure 5)
where it works reasonably well (amplitude of ray theory 25\%
higher than Born approximation kernels) and others where it is 
discrepant by a factor of two 
(\opencite{Birch07}, Figure 6).  
One case that works extremely well is the comparison of interior 
rotation determined from global modes and time--distance inversions
(\opencite{Giles00}, Figure 6.2) in the radius range $0.89<r/R_{\odot}<0.999$.
The difference between these cases seems to be that the ray theory works
better when the spatial variation of the perturbation is larger, that is,
much larger than the acoustic wavelength.
One way to check the ray
theory for a particular case is to compare with travel times computed from
Born kernels \cite{Birch07}.  
Another way to check is to perform simulations on a convectively--stabilized
solar model with
specified flow velocity perturbations and to propagate acoustic waves 
through the model \cite{Hanasoge07}.  
Travel times are computed from the simulations and
then compared with ray--theory calculations of the travel times using the
same flow velocity perturbations.  This is the type of checking of the ray
theory adopted here.

\begin{figure}
\centerline{\includegraphics[width=1.0\textwidth,clip=]{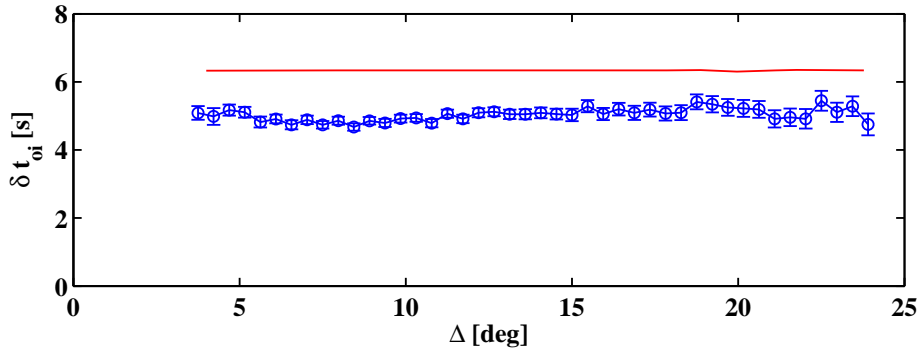}
           }
\caption{
Comparison of the center--annulus travel time differences $[\delta t_{\rm{oi}}]$
from the linear simulation (blue) with the travel--time difference
computed from the ray theory with the same flow perturbations (red).  
The error bars are computed from the
scatter far from the feature locations.  
No filtering has been done before the travel time
measurements.  The average travel--time difference in the range
$\Delta=10$\,--\,$24^\circ$ has been scaled to match the observationally 
determined mean
$5.1\secs$.  The same scaling factor is then used to scale the ray theory
results.  In the range $\Delta=10$\,--\,$24^\circ$, the ray theory predicts a
travel--time difference too large by 24\%.
}
\label{F-sim}
\end{figure}

A global simulation of wave propagation \cite{Hanasoge07b}
with $768 \times 384 $ grid points in longitude and latitude respectively 
is performed over a ten--hour period.
The flow perturbation consists
of Gaussian features with a constant direction of vertical flow centered at a
depth of $z_0=-2.30\Mm$ with $\sigma_z=0.82\Mm$ and with horizontal 
$\sigma_h=5.1\Mm$.
There are 500 of these features placed at random longitude--latitude pairs
at that depth $[z_0]$.
Center--annulus travel time differences are measured from the
simulation and averaged about the known locations.  Realization noise is
removed to first order by doing a similar simulation with no flow perturbation
but with the same source excitations \cite{Werne04, Hanasoge07} and subtracting
the resultant $\delta t_{\rm{oi}}$ to obtain noise--corrected results.

The flow model that is inserted into the simulation is used with 
Equation~(\ref{eq:tdif}) to derive ray--theory predictions of the travel
time differences.  The results of the ray--theory computations and the
travel time difference measurements from the simulation are shown in
Figure~\ref{F-sim}.  As the simulation only contains vertical flows, 
it is expected that there would be little or no variation of the travel--time 
differences with $\Delta$.  The ray theory does predict too large a 
travel--time difference by 24\%.  
This excess amplitude is similar to that found by \inlinecite{Birch07}.
Although the amplitude computed from the ray theory is not precise, 
the necessity of large vertical
flows to generate a travel--time difference of $5.1\secs$ at 
$\Delta=10$\,--\,$24^\circ$ is confirmed.  
The peak vertical flow for the normalized case is $338 \ms$.  
The model parameters are detailed in Table~\ref{T-models}.

\section{Data Reduction and Analysis}
\label{sec:danal}

\subsection{Reduction}
Dopplergrams from the {\it Helioseismic and Magnetic Imager} (HMI): 
\cite{Schou12} onboard the
{\it Solar Dynamics Observatiory} (SDO) spacecraft were analyzed for the present
work.  32 days (10 June\,--\, 11 July 2010) of Dopplergrams were used 
to derive
the final results, although a number of tests subsequently mentioned used
only the last three days, 9\,--\,11 July 2010.  This particular time period 
was used as the Sun was very quiet (sunspot number $RI\approx 15$ for these two 
months), and it was the final two--month continuous coverage period for
the {\it Michelson Doppler Imager} (MDI) instrument \cite{Scherrer95}.
It might be useful to compare the results from the two 
instruments, but only the HMI data was used for the present study.

Raw Dopplergrams have a nine--day averaged Dopplergram subtracted as well as
the spacecraft velocity which has a significant 24--hour component.
These corrected Dopplergrams are remapped onto a coordinate system with
equal spacing in latitude and longitude of $0.03^\circ$ covering a range
of $144^\circ$ in both latitude and longitude.  The remapping is achieved
using a bilinear interpolation.  The remapping in longitude is onto the
Carrington system at a central longitude that crosses central meridian at the
middle of the twelve--hour interval used.  Two twelve--hour intervals are used
for each day, covering the first and second halves of the day.
Each remapped image is Fourier--filtered and resampled at $0.24^\circ$ per
pixel.  Twelve--hour datacubes are constructed from these individual images
with $600\times600$ spatial points and 960 temporal points 
for the $45\s$ sampling.
This procedure works well for the large $\Delta$ emphasized in the present
study, but would need to be modified to study smaller $\Delta$.
Studying smaller $\Delta$ is deferred until future work.

The centers of the supergranules are located by a procedure used previously
\cite{Duvall10}.  The datacubes are spatio--temporally filtered
to pass just the solar {\it f}--mode oscillations.  Center--annulus travel time 
differences are computed \cite{Duvall00} for the distance range 
$0.48$\,--\,$1.02^\circ$.
By using the difference of inward--going times minus outward--going times,
the maps are equivalent to maps of the horizontal divergence.
The equivalence of these maps to maps of supergranules has been shown
before \cite{Duvall00,Duvall10}.
The travel--time difference maps are smoothed with a Gaussian of 
$\sigma=2.9\Mm$ to more easily determine the cell--center locations.
Local maxima of this signal are picked as the cell centers.  
Lists of the locations are made.  
The one of a pair with the smaller divergence that are closer together 
$23\Mm$ is rejected.
As it is desired 
to use center--annulus separations $\Delta <= 25^\circ$, locations  
close to the edge of the maps are not used.
On average, 930 features are located for each datacube with a total of
59549 for the 64 twelve--hour datasets.
Overlays of cell--center locations with the divergence maps are shown in
\cite{Duvall10}.

\subsection{Filtering}
\label{sec:filt}
Ray theory has proven invaluable in the development of time--distance 
helioseismology \cite{Duvall93}.  It led to the basic idea that signals
propagating along ray paths between surface locations would lead to 
correlations between the temporal signals from which travel times could
be inferred.  As acoustic, or {\it p}, modes with the same horizontal phase
speed $[\omega/k$: $\omega$ angular frequency and $k$ horizontal wavenumber$]$
travel along the same ray path to first order, it was natural to filter the data
in $\omega/k$ to isolate waves traveling to a certain depth.  This type
of filter was introduced by \inlinecite{Duvall93} with subsequent 
development by \inlinecite{Kosovichev97} and \inlinecite{Giles00}.
In the spatio--temporal power spectrum of solar oscillations 
($k$--$\omega$ 
diagram), a line from the origin has a constant phase speed $\omega/k$.
A range of phase speeds thus corresponds to a pie--shaped wedge.
Phase--speed filtering then correponds to multiplying the Fourier transform
of the input datacube by a pie--shaped wedge (appropriately modified for
3D) with some Gaussian tapering in phase speed \cite{Couvidat06}.

\begin{table}
\caption{
Nominal phase speed filter parameters.  The first column is an identification
number.  The second column is the range in distance $[\Delta]$.  
The third column is the mean distance $[\Delta_{\rm{mean}}]$.
The fourth
column is the central phase speed.  The fifth column is the full width at
half maximum (FWHM) of  the Gaussian phase speed filter.  For the tests in
Section~\ref{sec:filt}, filters 1\,--\,13 were used while for the final results
in Section~\ref{sec:gamma}, filters 1\,--\,14 were used. 
}
\label{T-filtering}
\begin{tabular}{rcrrr}
\hline
Filter  & $\Delta[^\circ]$ & $\Delta_{\rm{mean}}[^\circ]$ & $v_0[\kms]$ & $\delta v[\kms]$\\
\hline
1 & 1.56-2.28   & 1.92 & 32.0   & 13.2\\
2 & 2.28-3.00  & 2.64 & 41.8    & 4.2\\
3 & 3.00-3.96  & 3.48 & 46.9    & 6.0\\
4 & 3.96-4.92  & 4.44 & 54.7    & 9.5\\
5 & 4.92-6.12  & 5.52 & 64.5   & 10.1\\
6 & 6.12-7.80  & 6.96 & 75.1   & 11.1\\
7 & 7.56-9.24  & 8.40 & 84.3   & 10.4\\
8 & 9.12-10.7 &  9.84 & 93.5    & 9.4\\
9 & 10.92-12.60 & 11.76 & 104.4    & 9.7\\
10 & 12.36-14.52 & 13.44 & 114.2   & 12.5\\
11 & 14.28-16.44 & 15.36 & 125.2   & 12.3\\
12 & 17.16-19.32 & 18.24 & 141.9   & 12.6\\
13 & 19.08-22.20 & 20.64 & 156.0   & 18.5\\
14 & 22.08-24.00 & 23.04 & 170.4   & 11.8\\
\hline
\end{tabular}
\end{table}

The overriding principle of the use of the phase--speed filter is that
solar perturbations lead to refractive time changes for signals travelling
along a ray path.  That this model is deficient was shown nicely in the
work of \inlinecite{Couvidat06} in which an artificial signal was detected
with different amplitude depending on the width $[\delta v]$ of the 
pie--shaped wedge.  They concluded that the Born--approximation kernels
\cite{Birch00}, which are derived based on a single scattering from 
solar perturbations, are more appropriate.  A better way to think of the
problem is to consider waves impinging on a perturbation which are
subsequently scattered.  

\begin{figure}
\centerline{\includegraphics[width=1.0\textwidth,clip=]{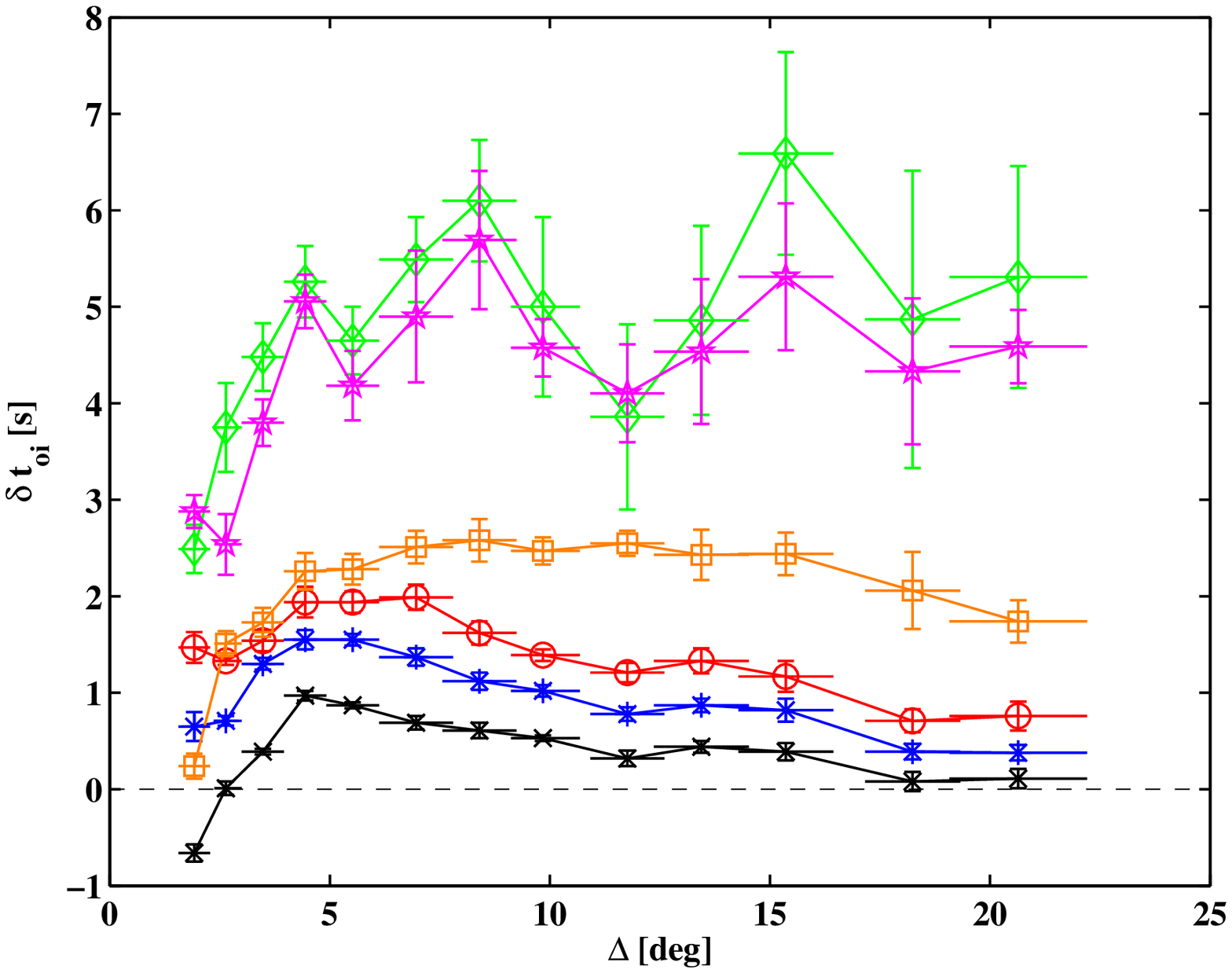}
           }
\caption{
Center--annulus travel time differences $[\delta t_{\rm{oi}}]$ 
averaged about the supergranule
centers for different filters for the thirteen $\Delta$ ranges for 
three days of HMI data.  From the bottom to top (on the right side), 
the black (x) points are for the the nominal phase--speed
filters.  The blue (*) points are for the phase--speed filters with the widths
doubled.  The red (o) points are for the width tripled.  The orange (square) points are
for a filter width that is half the nominal phase speed.  
The magenta (diamonds) curve is for the constant degree--width--filter 
(Section~\ref{sec:gamma}) with width $\Gamma_{\rm{\ell}}=400$.
The green (pentagrams) curve
is for no phase--speed filter.  In all cases the {\it f} mode is excluded via
filtering as is signal outside the frequency bandpass $1.5 < \nu<6\mHz$.
These results were obtained using the three days of data 9\,--\,11 July 2010.
}
\label{F-ph_noph}
\end{figure}

An initial analysis was performed using some nominal phase--speed
filters and distance ranges detailed in Table~\ref{T-filtering}.
Thirteen distance ranges (and hence
thirteen different filters) were applied to the six input datacubes for
the dates 9\,--\,11 July 2010.  
The overall distance range covered for this test is $3.0$\,--\,$22.2^\circ$. 
Center--annulus travel--time differences $[\delta t_{\rm{oi}}]$ 
were obtained for the thirteen 
distance ranges for each of the six datacubes using a standard 
surface--focusing time--distance analysis with the Gizon--Birch method of
extracting travel times from the correlations \cite{Gizon04}.
The travel times are averaged over the distance range and 
about the locations of the
supergranular centers and the signal at the 
average cell center location is extracted.
Uncertainties are estimated from the scatter 
of the six values.
The travel--time differences are plotted {\it versus} $\Delta$
in Figure~\ref{F-ph_noph} (black).  
The variation with $\Delta$ does not agree with the models
in Figure~\ref{F-model1}, but peaks near $\Delta=5^\circ$ and
decays approximately to zero near $\Delta=20^\circ$.  

There was some concern that this distance dependence might have something
to do with the filtering, so a case was analyzed with no phase--speed
filtering.  The {\it f} mode was excluded and there was a frequency bandpass 
filter transmitting $\nu=1.5$\,--\,$6\mHz$.  The results of that analysis are 
also shown in Figure~\ref{F-ph_noph} (green).  
This shows a behavior with $\Delta$ much closer
to the models in Figure~\ref{F-model1} but with considerably more noise
and a much larger signal close to five seconds.  It was concluded that the
phase speed filters were significantly degrading the signal.  To test
this further, the filters were broadened by the factors 2, 3, 4, 5 
with the results of factors 2 (blue) and 3 (red) also shown in 
Figure~\ref{F-ph_noph}.  
More signal is obtained with the larger--width
filters, but even with the factor of 5, only about 50\% of the signal
is obtained and there is still a decay of the signal towards larger
$\Delta$.

Because of the approximate constancy of the simulation
travel times $[\delta t_{\rm{oi}}]$ with $\Delta$ in 
Section~\ref{sec:sim} and the approximate
constancy with $\Delta$ for the no phase--speed filter case, it was concluded
that the decay of $\delta t_{\rm{oi}}$ toward larger $\Delta$ 
in the phase--speed filtered results is an artifact
of the filtering.  Filtering would seem to be desirable because of the
large difference in errors for the unfiltered and filtered cases.  
But a filter is needed that gives a more unbiased result at the 
large $\Delta$ in order that the results not be overly dependent on the
modeling.
A first attempt at this was to use a set of filters with the
full width at half maximum  (FWHM) equal to half the phase speed
\cite{Birch06}.  The results of these measurements are also shown in
Figure~\ref{F-ph_noph} (orange).  
The dependence on $\Delta$ is somewhat reduced but there is
still some decay at the largest $\Delta$.  

A different way to look at this problem is to consider a {\it p} mode with 
frequency $\nu$ and spherical harmonic degree $\ell$ 
impinging on a supergranule.
To the extent that the supergranule is static, the mode frequency $\nu$
will be maintained and the scattering will spread power in the spectrum away
from the nominal value of $\ell$.  The scattering should spread power
over a width in $\ell$ that depends on the spectrum of supergranulation,
which peaks near $\ell=120$.  For similar values of the phase speed (which
can also be characterized by $\nu/\ell$), the spread power should be 
over a similar width in $\ell$ \cite{Chou96,Woodard02}.  
A filter that is centered at a certain
phase speed but whose width is constant with $\nu$ and characterized by
the FWHM in $\ell$ ($\equiv \Gamma_{\ell}$) may be what is needed.
A value of $\Gamma_{\ell}=400$ should capture most of the supergranular signal.
Such a filter was implemented and applied to the thirteen distance ranges
used.  The results are also shown in Figure~\ref{F-ph_noph}.  
The approximate constancy
with $\Delta>5^\circ$ and the larger values than for the normal phase
speed filters suggests that this type of filtering is a useful concept.
The filter used has a flat top of width $\Gamma_{\ell}/2$ and is tapered 
to zero by a cosine bell that goes from 1 to 0 in $\Gamma_{\ell}/2$.

\begin{figure}
\centerline{\includegraphics[width=1.0\textwidth,clip=]{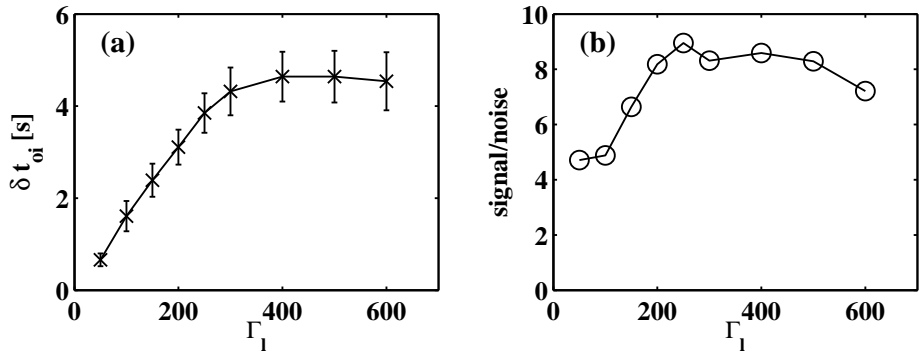}
           }
\caption{
(a) Travel--time difference $[\delta t_{\rm{oi}}]$ versus filter 
FWHM $[\Gamma_{\ell}]$.  The 
unfiltered case has $5.3\pm1.2\secs$.  (b) The travel time difference from (a)
divided by the size of the error bar from (a) {\it versus} the filter 
FWHM $\Gamma_{\ell}$.  The value for the unfiltered
case is 4.6.
}
\label{F-tt_and_sn}
\end{figure}

Some additional tests of this new filter concept were computed.  In 
Figure~\ref{F-tt_and_sn}(a)
is shown the peak supergranular signal for $\Delta=19.08$\,--\,$22.2^\circ$ 
as a function
of the filter width $\Gamma_\ell$. 
For small $\Gamma_\ell$, much of the supergranular signal is
removed, but as the width is increased more and more of the signal is
transmitted. There is clearly a variation in the size
of the error bars across Figure~\ref{F-tt_and_sn}(a).  
Another thing to examine is the ratio
of the signal in Figure~\ref{F-tt_and_sn}(a)  to the size of the error bar, 
which should give
a value of the signal to noise ratio (S/N) of the filter
\cite{Couvidat06}.  This is plotted
in Figure~\ref{F-tt_and_sn}(b).  
A rather broad peak is seen more or less centered on
$\Gamma_\ell=400$.  One noticeable aspect of Figure~\ref{F-tt_and_sn}(b) is 
that the unfiltered
case gives a worse S/N (4.6) than the filters near the peak.  
Also narrow filters
have reduced S/N.  Based on the flatness of the $\Gamma_\ell=400$ results
with $\Delta$ in Figure~\ref{F-ph_noph}, 
the extraction of essentially all the supergranular
signal in Figure~\ref{F-tt_and_sn}(a), and the value of the S/N in 
Figure~\ref{F-tt_and_sn}(b), $\Gamma_\ell=400$ 
has been chosen for further analysis.

\subsection{$\Gamma_\ell=400$ Analysis}
\label{sec:gamma}
The 64 twelve--hour datacubes were analyzed with the fourteen phase--speed 
filters with $\Gamma_\ell=400$.  Center--annulus cross correlations were 
computed for the various distance ranges of Table 1.  For checking purposes,
it would be useful to be able to compute travel times using both the
Gabor wavelet method \cite{Duvall97} and the Gizon--Birch method
\cite{Gizon04}.  However, for the large distances in the present study
($\Delta<24^\circ$), individual twelve--hour cross correlations have 
insufficient signal--to--noise ratio to enable the Gabor wavelet fitting.
It was decided to average the temporal correlations spatially about the 
supergranular centers.  The averaging of the 930 (on average) correlations 
yields a high signal--to--noise ratio correlation that is amenable to the
Gabor wavelet fitting at the largest distances $[\Delta]$.  Averaging the 
resultant $\delta t_{\rm{oi}}$ 
for the 64 twelve--hour datacubes enables the use of the envelope--time 
differences $[t_{\rm{env}}]$ as well as the normal phase times 
$[t_{\rm{ph}}]$.
Although most theoretical work applies to phase--time differences,
it will be useful to examine the envelope--time differences as they may have
independent information.

\begin{figure}
\centerline{\includegraphics[width=1.0\textwidth,clip=]{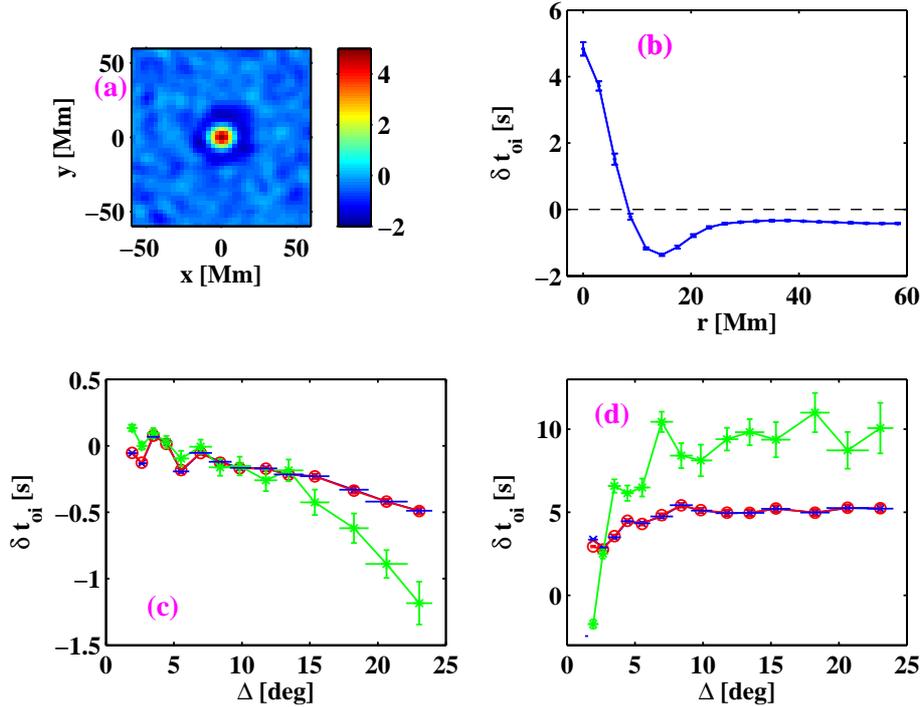} }
\caption{
(a) Center--annulus travel--time difference averaged about supergranule 
centers for the range of $\Delta=22.08$\,--\,$24^\circ$.  The scale of the 
colorbar at right is in seconds.  Only the central point corresponding
to the supergranular center is used in the present study.
(b) Azimuthal average of (a).  Note the offset at large radii.
This offset is believed to be an artifact which needs to be removed from
the results.
(c) The offset at $r=58\Mm$ for the different travel--time definitions
{\it versus} $\Delta$.
Blue is for the Gabor wavelet phase time differences.  Red is for the
Gizon--Birch phase time differences and the green is for the Gabor
wavelet envelope time differences.
(d) The resultant travel time differences averaged for the
64 12--hour datacubes corrected for the offset in (c).  The colors are the
same as in (c).
}
\label{F-errplot}
\end{figure}

It was found previously that there is an offset from zero 
(of unknown origin) of the
center--annulus travel--time differences \cite{Duvall00}.  For the small
$\Delta$ of that study, the value of the offset was $0.16\pm0.02\secs$.  
This offset is assumed unrelated to the flows to be measured and 
needs to be removed from computed travel times.  
To measure it for the present study, the 
average maps about the supergranule centers for the 64 12--hour datacubes
are computed for the different distance ranges.  An example is shown for
the largest range of $\Delta=22.08$\,--\,$24^\circ$ in Figure~\ref{F-errplot}(a).
An azimuthal average of Figure~\ref{F-errplot}(a) is computed and the
result is shown in Figure~\ref{F-errplot}(b).  At large radii $[r]$, the
the signal is approximately constant.  This constant, which is assumed to
be the offset that needs to be subtracted from the results, is plotted
for the various $\Delta$ and travel--time methods in 
Figure~\ref{F-errplot}(c).  The Gizon--Birch and Gabor wavelet phase times
yield almost exactly the same results, which is expected for the
quiet Sun and the use of the same time window for fitting
(20 minutes).
The travel--time differences corrected for this offset are plotted in
Figure~\ref{F-errplot}(d).  Again, the Gizon--Birch and Gabor wavelet phase
times yield almost precisely the same results
with an average of $5.1\pm0.1\secs$ on the range $\Delta=10$\,--\,$24^\circ$.  
The Gabor wavelet 
envelope times yield a considerbly larger value of $9.7\pm0.3\secs$ on
$\Delta=10$\,--\,$24^\circ$.
In general the group velocity, $\partial\omega/\partial k$ is
about one half of the phase speed $\omega/k$ for {\it p} modes, possibly 
leading to the larger travel time.  The systematic error from 
Figure~\ref{F-errplot}(c) is generally less the 10\% of the 
signal shown in Figure~\ref{F-errplot}(d).  

\begin{figure}
\centerline{\includegraphics[width=1.0\textwidth,clip=]{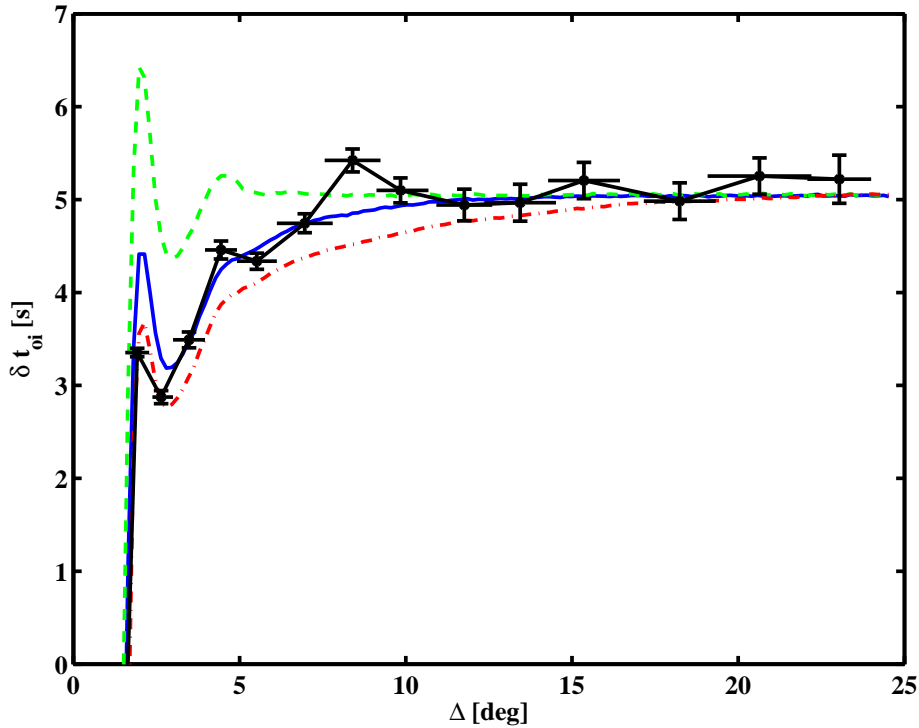} }
\caption{
Comparison of the three ray models from Figure~\ref{F-model1}
with the HMI results (black with symbols and errors) 
of the $\Gamma_\ell=400$ filtering and the 
Gabor--wavelet phase--speed time differences from Figure~\ref{F-errplot}(d). 
These models
are the sum of vertical and horizontal signals for the three Gaussian
vertical flows peaking at $z_0=-3.45\Mm$ (red dot--dashed), 
$z_0=-2.3\Mm$ (blue solid), and
$z_0=-1.15\Mm$ (green dashed) specified in Table~\ref{T-models}.  
}
\label{F-cmpplot}
\end{figure}

The Gabor--wavelet phase--time differences are plotted in 
Figure~\ref{F-cmpplot} {\it versus}
the three ray models of Figure~\ref{F-model1}.
The increase of the observed times for $\Delta=2.5$\,--\,$8^\circ$ may be
the effect of the horizontal flow.  
The agreement is a little better for the $z_0=-2.3\Mm$ model.

\section{Discussion}
\label{sec:discuss}
The main observational result of this article is that in the distance
range $\Delta=10$\,--\,$24^\circ$ that the mean travel--time difference
$[\delta t_{\rm{oi}}]$ 
at the center of the average supergranule 
is $5.1\pm0.1\secs$.  How secure is this result?
There are no other results to compare with in this distance range.
Also, the method of averaging over many supergranules is uncommon
\cite{Birch06,Duvall10,Svanda11}.  The largest distance $[\Delta]$ used
previously is $5^\circ$ by \inlinecite{Zhao03}.  
The phase--speed filter used in that
study would have reduced the travel--time difference 
$[\delta t_{\rm{oi}}]$ by a factor of five and so it is difficult to compare.
In addition, only inversions were presented and not raw travel--time
differences.  

One of the most interesting aspects of the present results is the large
factor ($>20$) between the photospheric vertical flow and the peak vertical 
flow.  This increase of vertical flow then also requires an increase of
the horizontal flow from the photosphere to the peak of a factor
between 3.7 and 8.3 from the simple Gaussian models g1\,--\,g3.
Some simple tests suggest that this horizontal velocity should be
detectible by a quadrant time--distance analysis, which will be left
for future work.  
Some previous analyses
have detected some larger flows below the surface than at the surface
\cite{Duvall97}.  Possibly the subsurface flow is highly variable 
which has made average properties difficult to ascertain from 
inversions of individual realizations.

These results confirm that supergranulation is a shallow phenomenon, with
the flow peaking within a few Mm of the photosphere.  This would tend to
support models in which supergranulation is related to a near--surface
phenomenon such as granulation \cite{Rast03}.  
The supergranular wave measurements
\cite{Gizon03,Schou03}, and in particular the variation with latitude
of the anisotropy of wave amplitude, would suggest a connection with 
rotation.  The near--surface shear layer has been modeled as a possible exciter
of supergranulation \cite{Green06}.  \inlinecite{Hathaway82} found that
supergranulation as a convective phenomenon generates a near--surface shear
layer.
A difficulty with the near--surface shear layer as supergranular exciter
is that recent simulations of \inlinecite{Stein09}
develop a near--surface shear layer but no hint of excess power at
supergranular scales is seen.



\begin{acks}
The data used here are courtesy of NASA/SDO and the HMI Science Team.  We
thank the HMI team members for their hard work.  This work is supported by
NASA SDO and the NASA SDO Science Center program through grant
SCEX22011D awarded to NASA GSFC.
S. M. H. acknowledges funding from NASA grant  NNX11AB63G. 
\end{acks}

\end{article} 

\end{document}